\documentclass[aps,prx, reprint,longbibliography, nofootinbib,superscriptaddress]{revtex4-1}

\usepackage{subfigure}
\usepackage{graphicx,amsmath,amssymb,color,amsfonts,bm,physics}
\usepackage{mathtools,lipsum}
\usepackage[dvipsnames]{xcolor}
\usepackage[misc]{ifsym}
\usepackage{quantikz}
\usepackage{algorithm}
\usepackage{algpseudocode}
\usepackage[colorlinks=true]{hyperref}  
\hypersetup{
    bookmarks=true,         
    unicode=false,          
    pdftoolbar=true,        
    pdfmenubar=true,        
    pdffitwindow=false,     
    pdfstartview={FitH},    
    pdftitle={},    
    pdfauthor={},     
    pdfsubject={},   
    pdfcreator={},   
    pdfproducer={}, 
    pdfkeywords={} {} {}, 
    pdfnewwindow=true,      
    colorlinks=true,       
    linkcolor=blue, 
    citecolor=blue,        
    filecolor=magenta,      
    urlcolor=blue,           
        breaklinks=true
} 

\newcommand{\be}{\begin{equation}}
\newcommand{\ee}{\end{equation}}
\newcommand{\bea}{\begin{eqnarray}}
\newcommand{\eea}{\end{eqnarray}}

\begin{document}

\title{Quantum state randomization constrained by non-Abelian symmetries}

\author{Yuhan Wu and Joaquin F. Rodriguez-Nieva}
\affiliation{Department of Physics \& Astronomy, Texas A\&M University, College Station, TX 77843}


\begin{abstract}

The emergence of randomness from unitary quantum dynamics is a central problem across diverse disciplines, ranging from the foundations of statistical mechanics to quantum algorithms and quantum computation. Physical systems are invariably subject to constraints---from simple scalar symmetries to more complex non-Abelian ones---that restrict the accessible regions of Hilbert space and obstruct the generation of pure random states. In this work, we show that for systems with noncommuting symmetries such as SU(2), the degree of Haar-like randomization achievable under unitary dynamics is strongly constrained by experimental limitations on state initialization, in particular low-entanglement initial states, rather than by the symmetry-constrained dynamics themselves. Specifically, we show that time-evolved states can, in principle, reproduce Haar-like behavior at the level of finite statistical moments (i.e., those accessible under realistic experimental conditions with a finite number of state copies) provided that the initial state matches the corresponding moments of the conserved operators in the Haar ensemble. However, for the unentangled initial states commonly used in programmable quantum systems, this condition cannot be satisfied. Consequently, even at asymptotically long times in strongly quantum-chaotic regimes, late-time states remain distinguishable from Haar-random states in probes such as entanglement entropy, with deviations from Haar behavior that remain finite with increasing system size. We quantify the maximal entanglement entropy achievable and identify the unentangled initial conditions that yield the most entropic late-time states. Our results show that the combination of non-Abelian symmetry structure and experimental constraints on state preparation can strongly limit the degree of Haar-like randomization achievable at late times.

\end{abstract}



\maketitle

\section{Introduction}

Understanding how randomness emerges through quantum-chaotic evolution is central both to the foundations of statistical mechanics in isolated quantum systems~\cite{1991PRL_Deutsch,1999JPA_Srednicki,2008Nature_Rigol,2011RMP_Quantumthermalization_review,2016RPP_thermalizationreview} and to many applications in quantum information science, including benchmarking~\cite{2023Nature_emergentdesign,2023PRL_chaosbenchmarking}, tomography~\cite{2020SIAM_shadowtomography,2020NatPhys_shadowtomography}, sensing~\cite{Fiderer2018,doi:10.1126/science.adg9500,2025PRL_ScramblingSensing}, and demonstrations of quantum computational advantage~\cite{2019Nature_quantumsuppremacy,2021Science_googlescrambling}. The standard expectation is that an initially unentangled state evolving under a generic quantum-chaotic Hamiltonian---away from fine-tuned limits such as integrable limits---becomes effectively featureless at late times. This expectation is supported by decades of theoretical and experimental work on quantum thermalization, and holds broadly when dynamics is probed through time-averaged local observables, thereby justifying the use of thermal ensembles to describe the statistical behavior of quantum matter.

In a stricter sense, however, realistic many-body systems have physical structure---from spatial locality to global symmetries---that strongly constrains their exploration of Hilbert space and prevents arbitrary states from being dynamically accessible. This already occurs in the presence of a simple U(1) symmetry, and becomes even more restrictive for SU(2). For example, generic quantum circuits comprised of gates that preserve U(1) or SU(2) symmetry are non-universal~\cite{Marvian2022}, and therefore cannot generate arbitrary quantum states. Similar limitations apply to Hamiltonian dynamics, where the overlap of an initial state with the energy eigenbasis is likewise constrained~\cite{2024PRX_maximumentropyprinciple,2025PRB_latetimeensemble}.

The apparent tension between the use of thermal ensembles in statistical mechanics and the impossibility of exploring arbitrary states in the presence of symmetries can be resolved by asking whether randomness emerges at the level of finite statistical moments~\cite{Harrow2009,HunterJones2019}. From this perspective, time-evolved states are effectively random insofar as they reproduce the finite-order moments of random-state ensembles, with the first moment capturing the thermal behavior described by conventional statistical mechanics and higher moments capturing state-to-state fluctuations beyond thermalization. This viewpoint is also operationally natural, since experiments have access only to a finite number of copies of a quantum state and can therefore probe it with finite resolution.

This finite-resolution perspective has already led to several notable insights. In systems constrained only by spatial locality, with no additional symmetries, Haar-like statistics can be generated in polynomial time or circuit depth, long before the exponentially long time required to explore the full Hilbert space~\cite{brandao2012local,chen2024incompressibility}. Moreover, in the presence of a U(1) symmetry, time-evolved states can still reproduce the same finite-order moments as Haar-random states, provided the initial state matches the corresponding moments of the conserved charge in the Haar ensemble~\cite{2025PRB_latetimeensemble}. Thus, although symmetry-constrained dynamics never explores the full Hilbert space, it can nonetheless exhibit effective randomness at the level of finite statistical moments---and can do so on polynomial timescales~\cite{2025arxiv_randomizationtimes}.

In this work, we ask how much randomness can be generated by quantum chaotic dynamics in the presence of noncommuting charges, focusing on SU(2) symmetry. In particular, we ask whether initially unentangled states evolving under SU(2)-symmetric dynamics can reproduce the statistics as Haar-random states moment by moment, with a particular focus on entanglement entropy as a sensitive probe of randomization. 

These questions are particularly timely given the ubiquity of non-Abelian symmetries across diverse physical settings, from complex nuclei and atomic systems~\cite{Bauer2023} to Heisenberg models in condensed matter~\cite{2020Nature_jepsen,2022Science_kpzexp} and non-Abelian gauge theories~\cite{Tagliacozzo2013}. Despite their prevalence, the dynamics of systems with noncommuting charges remain far less explored than their commuting counterparts. As recently shown in a variety of contexts, noncommuting symmetries can give rise to qualitatively new phenomena, including anomalous spin transport with KPZ scaling~\cite{2020Nature_jepsen,2022Science_kpzexp} (yet distinct scaling in higher-order fluctuations~\cite{2024Science_googlekpz}), universal prethermal dynamics after a quench~\cite{2020PRL_rodrigueznieva,2022PNAS_rodrigueznieva}, 
equilibration to non-Abelian thermal states~\cite{2023PRX_nonabelianexp,PRXQuantum.5.030355}, tensions with conventional assumptions underlying the eigenstate thermalization hypothesis (ETH)~\cite{2023NatRev_noncommuting,2023PRE_su2eth}, and distinct entanglement structure  relative to systems with commuting conserved charges~\cite{2023PRB_su2ee,2024SciPost_SU2Bianchi}. As we will see, we will find equally rich structures emerging from many-body quantum dynamics with SU(2) symmetry when one probes finer-grained diagnostics of randomization beyond thermal ensembles through the lens of entanglement.  

In addition, questions addressing many-body quantum dynamics beyond thermalization are highly relevant to modern experiments on programmable quantum platforms, including superconducting qubits~\cite{2019Nature_quantumsuppremacy,2024Science_googlekpz,2021Science_googlescrambling}, Rydberg atoms~\cite{2017Nature_51atomsimulator,2021Nature_Rydberg256,2021Nature_RydbertAFM}, and trapped ions~\cite{2017Nature_monroe,2021RMP_trappedions}. These systems provide access to extremely detailed statistical information about quantum states through microscopically resolved measurements of individual degrees of freedom, together with precise and highly repeatable unitary control. Such capabilities make it possible to measure not only expectation values of local observables, but also finer statistical properties of many-body states, including full counting statistics~\cite{2023Nature_emergentdesign,2023PRL_fullcountingstatistics,2024PRB_fullcountingstatistics} and subsystem entanglement entropies~\cite{2019Science_EEmbl,2018PRL_zollerrenyientropy,2019Science_renyiions,2023NatRev_randommeasurement}, both of which are sensitive to the fine-grained structure of quantum dynamics.

As a consequence of these new experimental capabilities, the theoretical focus has shifted from asking only whether a quantum many-body system thermalizes to asking what finer statistical structure remains visible in experimentally accessible observables. This has motivated a broad effort to characterize many-body dynamics beyond the coarse-grained predictions of thermal ensembles and random-matrix theory: for example, by identifying how spatial locality constrains the structure of quantum states~\cite{Huang2021,2022PRE_deviationsfromETH,2024PRX_quantifyingchaos,2025PRL_EEU1}, how such structure appears in level statistics and in correlations between matrix elements of local observables~\cite{2019PRL_chalker,2019PRE_kurchan,2020PRE_beyondETH,Garratt2021,2022PRL_dymarsky,2022PRA_Ranard,2021PRE_eth_otocs}, and how to construct statistical ensembles that capture experimentally resolvable features of quantum-state distributions beyond thermalization~\cite{2023PRX_cotleremergentdesign,2022PRL_exactdesigns,2022Quantum_designs,2023PRL_completeergodicity,HunterJones2019,2020PRA_higherorderETH,PhysRevX.15.011031}.

\begin{figure}
    \centering
    \includegraphics[width=\linewidth]{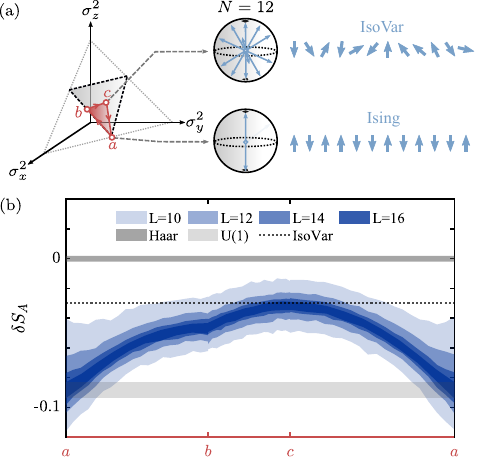}
    \caption{(a) The late-time behavior of unentangled initial states with zero magnetization is governed 
    by the spin variances of the initial state, which are constrained by the condition $\sigma_x^2+\sigma_y^2+\sigma_z^2 = L/2$. This defines a region of physically allowed states in $(\sigma_x,\sigma_y,\sigma_z)$ space (shaded area bounded by dashed lines).  We focus on initial states along the high symmetry lines, with high-symmetry points denoted by $a$, $b$, and $c$: point $a$ corresponds to a product state with $\sigma_z = 0$ (i.e., an Ising state with spins aligned in the $z$ direction), whereas point $c$ corresponds to a product state with 
    all spins uniformly distributed on the Bloch sphere, referred to as `IsoVar'. 
    (b) Distribution of half-system entanglement entropy (EE) at late times as a function of system size $L$ and spin variance of the initial conditions along the $a$--$b$--$c$ line. The data is shown relative to the EE of pure (Haar) random states, $\delta S_A = \langle S_A\rangle - \langle S_A\rangle_{\rm Haar}$, with shaded regions indicating the standard deviation of the EE. For comparison, we indicate with dark (light) gray areas the distribution of EE of the Haar random states (random states constrained by a U(1) charge). The EE reaches a maximum value at point $c$ corresponding to the IsoVar initial condition, which is lower than the maximal EE of Haar random states, and the difference remains finite over the system sizes studied and is consistent with a finite thermodynamic-limit offset.}
    \label{fig:schematics}    
\end{figure}

\subsection{Summary of key results}

The main results of this work are summarized in Fig.~\ref{fig:schematics}. Our analysis begins by showing that, even when states are constrained to preserve symmetries, the resulting constrained ensemble can still exhibit the same statistics of Haar random states when both ensembles are compared moment by moment, up to some finite order $k$. For this to occur, the moments of all conserved operators $S_\alpha$ must match those of the Haar ensemble, i.e.,
\be
\begin{aligned}
\bar S_\alpha &= \langle \psi_0 | S_\alpha | \psi_0 \rangle 
               = \frac{1}{D}{\rm Tr}[S_\alpha] = 0, \\
\sigma_\alpha^2 & = \langle \psi_0 | S_\alpha^2 | \psi_0 \rangle 
               = \frac{1}{D}{\rm Tr}[S_\alpha^2] = LS^2,
\label{eq:condition}          
\end{aligned}
\ee
and similarly for higher moments and for all three spin components $\alpha = x,y,z$. The condition~(\ref{eq:condition}) needs to hold for a single element of the constrained ensemble (e.g., the initial state $|\psi_0\rangle$) and it then automatically holds for all other states in the ensemble since the dynamics preserves not only total magnetization but also all of its moments. 
If (\ref{eq:condition}) is satisfied, distinguishing the constrained ensemble from Haar-random states would require an exponential number of copies to resolve the exponentially small differences between the statistics of the two ensembles. These results generalize our earlier findings for U(1)-conserving systems to the case of non-Abelian symmetries~\cite{2025PRB_latetimeensemble}.

Second, we show that for the broad and experimentally relevant class of {\it unentangled} initial states, the condition (\ref{eq:condition}) cannot be satisfied. This can be understood most simply by noting that the variances of the three spin components cannot be made arbitrarily large: for a system comprised of $L$ spin-1/2 degrees of freedom, the spin variance of any product state necessarily obeys the identity $\sigma_x^2+\sigma_y^2+\sigma_z^2 = L/2$, which is parametrically smaller than the corresponding spin variance of a Haar-random state, $\sigma_x^2+\sigma_y^2+\sigma_z^2 = 3L/4$ (as $\sigma_x^2 = \sigma_y^2 = \sigma_z^2 = L/4)$. As illustrated in Fig. 1, for all unentangled initial states studied, the late-time entanglement entropy (EE) remains below the Page entropy characteristic of Haar-random states, consistent with a finite $O(1)$ shift that persists in the thermodynamic limit.

Third, we show that the late-time entanglement statistics of initially unentangled states with zero total magnetization are primarily governed by the spin variances along each direction, and are otherwise largely insensitive to the microscopic realization of those variances. In particular, the condition $\sigma_x^2 + \sigma_y^2 + \sigma_z^2 = L/2$ defines a two-dimensional plane that parametrizes families of initial states [panel (a) in Fig.~\ref{fig:schematics}], and we find that all states with the same $(\sigma_x,\sigma_y,\sigma_z)$ values have the same mean and state-to-state fluctuations of EE, up to numerical accuracy. 

Fourth, by systematically exploring the full space of unentangled initial conditions, we identify two extreme limits. At one end, when a single charge is fully resolved (point $a$ in Fig.~\ref{fig:schematics}a) while the variances in the remaining directions are maximized (e.g., $\sigma_z=0$, $\sigma_x^2 = \sigma_y^2 = L/2$), we recover the same O(1) corrections to the EE known from U(1)-conserving systems~\cite{2017PRL_EEchaotic,2019PRD_BianchiDona,2022PRX_eereview}: the system behaves as if having a single scalar charge, since information about magnetization along the $x$ and $y$ directions is effectively erased during evolution. 
In contrast, when fluctuations are distributed as uniformly as possible among the three spin components---so that each charge has equal but submaximal variance (point $c$ in Fig.~\ref{fig:schematics}a)---we observe $O(1)$ deviations from maximal entanglement entropy that persist in the thermodynamic limit, but are minimal among all admissible initial states. We propose and numerically validate a quantitative form for the $O(1)$ corrections to the maximally allowed EE, and show remarkably good agreement with numerics for all accessible system and subsystem sizes and across the full range of initial conditions. 

\section{Non-abelian constraints on dynamics}
\label{sec:theory}

In this section, we describe the three main theoretical results of this work, namely that: 
(i) States constrained to preserve symmetries $S_\alpha$ can still reproduce the moment-by-moment statistical behavior of Haar-random states up to some finite resolution (i.e., polynomial in system size), provided that the initial state $|\psi_0\rangle$ matches the full distribution of $S_\alpha$ in the Haar ensemble [Eq.~(\ref{eq:condition})]; (ii) For the special class of unentangled initial states, which are states commonly used as initial conditions in experiments, the previous condition (i) cannot be satisfied in the presence of non-Abelian symmetries; (iii) There exists a bound, which we quantify below, on the maximum EE that can be generated at late times when the initial state is unentangled. All these results will be supported by extensive numerical calculations in random quantum circuits (RQCs) (Sec.~\ref{sec:RQCs}) and in spin Hamiltonians with SU(2) symmetry (Sec.~\ref{sec:Ham}). 

\subsection{Random states constrained by non-Abelian symmetries}
\label{sec:contrainedensemble}

We first show that symmetry-constrained states can nonetheless display Haar-level statistics despite exploring only a small subspace of Hilbert space. To this end, we construct a constrained ensemble that mimics time-evolved states: the average value and all moments of the conserved charges---fixed by the initial condition---are preserved, while the remaining information about the state is randomized by quantum-chaotic dynamics. We show that this ensemble reproduces both the mean behavior and state-to-state fluctuations of the Haar ensemble up to exponentially small corrections. In particular, we quantify distinguishability between ensembles through the trace distance between ensembles, which determines the maximum success probability of distinguishing them under the optimal (not necessarily local) measurement. The calculation follows the spirit of our previous work~\cite{2025PRB_latetimeensemble} on U(1)-conserving systems, here generalized to the SU(2) case by incrementally incorporating additional symmetry structure.

To illustrate this result within a simple toy model, we define the ensemble
\be
\Phi = \left\{ |\Phi_{i}\rangle = \sum_{S,M} \sqrt{\frac{D_S}{D}} |\phi^{(i)}_{S,M}\rangle, \quad i = 1,2,\ldots \right\},
\label{eq:constrainedensemble}
\ee
for $L$ spin-1/2 degrees of freedom, where $D = 2^L$ denotes the total Hilbert space dimension, $D_S = {L \choose L/2-S} - {L \choose L/2-S-1}$ denotes the Hilbert-space dimension of each symmetry sector with total spin $S$ associated with the $S^2 = S_x^2+S_y^2+S_z^2$ operator. Each of the states $|\phi^{(i)}_{S,M}\rangle = \sum_{\alpha=1}^{D_S}\phi_{S,M,\alpha}^{(i)}|S,M,\alpha\rangle $ is a pure random state drawn independently within each sector labeled by spin $S$ and magnetization $M$ in the $z$ direction, with $|S,M,\alpha\rangle$ the basis elements of that sector and $\sum_{\alpha=1}^{D_S} |\phi_{S,M}^{(i)}|^2 = 1$. Here we adopt the convention that ${N \choose k}=0$ for $k<0$. Importantly, the ensemble $\Phi$ distributes weight across the different symmetry sectors in proportion to their Hilbert-space dimensions $D_S$, and this weight remains fixed for all elements of the ensemble. 

By construction, the ensemble $\Phi$ samples only a small fraction of Hilbert space. For instance, states with staggered magnetization—despite having zero total magnetization—are not included in $\Phi$, as they have well-resolved $M$ and reside only in the $M=L/2$ sector.

In addition, $\Phi$ mimics key constraints imposed by unitary evolution under SU(2). In particular, each state in the ensemble has zero total magnetization, $\langle \Phi_i | S_z | \Phi_i \rangle = \frac{1}{D}{\rm Tr}[S_z] = 0$, as well as identical higher moments, $\langle \Phi_i | S_z^2 | \Phi_i \rangle = \frac{1}{D}{\rm Tr}[S_z^2] = L$, \ldots, as required by unitary dynamics once the initial state fixes the distribution of conserved charges. The same holds for the total-spin operator, $\langle \Phi_i | S^2 | \Phi_i\rangle = \frac{1}{D}\operatorname{Tr}[S^2]$, and for all higher statistical moments of $S^2$. 

The key distinction between $\Phi$ and the behavior of time-evolved states is that $\Phi$ satisfies $\langle S_{x,y} \rangle_{\Phi} = \sum_i \langle \Phi_i | S_{x,y} | \Phi_i\rangle =\frac{1}{D}{\rm Tr}[S_{x,y}] = 0$ only {\it on average} but not at the level of individual states $i$, as in general $\langle \Phi_i | S_x | \Phi_i \rangle \neq 0$. This contrasts with SU(2)-conserving dynamics, where the expectation values of ${S}_x$ and ${S}_y$ are fixed from the onset of dynamics by the initial condition and remain fixed for {all} time-evolved states. Relaxing the constraint from being enforced exactly for each $|\Phi_i\rangle$ to being satisfied only statistically greatly simplifies the analysis and renders the calculations more transparent and analytically tractable. This already illustrates the central point: as additional constraints are incorporated---progressing from conservation of $S_z$ (as considered in our previous work~\cite{2025PRB_latetimeensemble}), to the inclusion of $S^2$, and eventually to the full SU(2) structure including $S_{x,y}$---the ensemble remains exponentially close to the Haar ensemble for finite-moment probes, including nonlocal observables. Appendix~\ref{app:fullSU2} shows numerically that enforcing the stronger constraint $\langle \Phi_i | S_{x,y} | \Phi_i\rangle = 0$ at the level of individual states leads to the same qualitative conclusions for the observables considered here.

We now compute the statistical behavior of states drawn from $\Phi$ moment by moment and quantify their deviation from the ensemble of pure random states. As a reminder, when considering pure random states $|\Psi_i\rangle = \sum_{\alpha=1}^{D}\psi_\alpha^{(i)}|\alpha\rangle$ in a Hilbert space of dimension $D$, the first moment of the distribution is defined as $\rho_{\rm Haar}^{(k=1)} =\sum_i p_i|\Psi_i\rangle\langle\Psi_i|$, where $p_i$ is the probability of drawing sample $i$. Here we assume that all states are equally likely to occur, thus $p_i = 1/N_{\rm s}$ and $N_{\rm s}\rightarrow\infty$. The matrix elements $[\rho_{\rm Haar}^{(k=1)}]_{\alpha,\alpha'} $ are 
\be
\langle \psi_{\alpha}\bar{\psi}_{\alpha'}\rangle_{\rm Haar} = \frac{\delta_{\alpha,\alpha'}}{D}.
\label{eq:Haarfirstmoment}
\ee
Essentially this equality highlights that Haar random states have uncorrelated wavefunction amplitudes with zero mean and variance $1/D$. 

Similarly, the second moment of the Haar ensemble is quantified by $\rho_{\rm Haar}^{(k=2)} = \sum_i p_i |\Psi_i\rangle\langle\Psi_i|\otimes|\Psi_i\rangle\langle\Psi_i|$, where the matrix elements of $[\rho_{\rm Haar}^{(2)}]_{\alpha_1\alpha_2,\alpha_1'\alpha_2'}$ are:
\be
\langle \psi_{\alpha_1}\bar{\psi}_{\alpha_1'}\psi_{\alpha_2}\bar{\psi}_{\alpha_2'}\rangle_{\rm Haar} = \frac{\delta_{\alpha_1,\alpha_1'}\delta_{\alpha_2,\alpha_2'}+\delta_{\alpha_1,\alpha_2'}\delta_{\alpha_2,\alpha_1'}}{D(D+1)}.
\label{eq:Haarsecondmoment}
\ee
This result can be obtained from taking ensemble average of the square of the normalization condition $\sum_{\alpha,\alpha'}\langle \psi_{\alpha}\bar{\psi}_{\alpha}\psi_{\alpha'}\bar{\psi}_{\alpha'}\rangle =1$, which contains $D(D-1)$ equivalent terms with $\alpha\neq \alpha'$, and $D$ terms with $\alpha = \alpha'$ which contribute twice to the summation (as there are two ways to contract the Gaussian variables). This gives rise to a total of $D(D+1)$ equivalent terms that sum to 1.

Returning to $\Phi$ in Eq.~(\ref{eq:constrainedensemble}), we simplify the notation by introducing the composite index $Q=(S,M)$ to label a symmetry sector with quantum numbers $S$ and $M$, and $\alpha$ labeling the basis states within each sector $Q$, $\alpha = 1,2,\ldots,D_S$. 
The average behavior of $\Phi$ is encoded in the density matrix $\rho^{(k=1)}_{\Phi} = \sum_ip_i|\Phi_i\rangle\langle\Phi_i|$, with matrix elements $[\rho_\Phi^{(k=1)}]_{Q_1\alpha_1,Q_2\alpha_2}=\langle \Phi_{Q_1\alpha_1} \bar\Phi_{Q_2\alpha_2} \rangle_\Phi$:
\be
\langle \Phi_{Q_1\alpha_1} \bar\Phi_{Q_2\alpha_2} \rangle = \frac{\sqrt{D_{1}D_{2}}}{D} \frac{\langle\phi_{1}\bar\phi_{2}\rangle}{D_{1}} = \frac{\delta_{1,2}}{D},
\ee
where we used the short-hand notations $\phi_i = \phi_{Q_i,\alpha_i}$, $D_i = D_{S_i}$ and $\delta_{1,2} = \delta_{Q_1,Q_2}\delta_{\alpha_1,\alpha_2}$, and that $|\phi_{Q,\alpha}\rangle$ are normalized random vectors within each sector $Q=(S,M)$ ($\langle \phi_{Q_1\alpha_1} \bar\phi_{Q_2\alpha_2} \rangle = \delta_{Q_1 Q_2}\delta_{\alpha_1\alpha_2}/D_{S_1}$). Hence, the first moment of the ensemble in Eq.~(\ref{eq:constrainedensemble}) coincides exactly with the Haar first moment. At the level of average expectation values, the constrained ensemble is therefore indistinguishable from the Haar ensemble, despite exploring only a symmetry-restricted subspace of the Hilbert space.

The second moment of the ensemble is encoded in the density matrix $\rho^{(k=2)}_{\Phi} = \sum_ip_i|\Phi_i\rangle\langle\Phi_i|\otimes|\Phi_i\rangle\langle\Phi_i|$, with matrix elements $[\rho_\Phi^{(k=2)}]_{Q_1\alpha_1Q_2\alpha_2,Q_1'\alpha_1'Q_2'\alpha_2'}$ given by:
\begin{widetext}
\begin{equation}
\begin{aligned}
\big\langle \Phi_{Q_1\alpha_1}\bar\Phi_{Q_1'\alpha_1'}\Phi_{Q_2\alpha_2}\bar\Phi_{Q_2'\alpha_2'} \big\rangle_{\Phi}
&={\frac{D_1D_2}{D^2}}\;\langle\phi_1\bar\phi_{1'}\phi_2\bar\phi_{2'}\rangle = 
\begin{cases}
\displaystyle 
\frac{\delta_{1,1'}\delta_{2,2'}+\delta_{1,2'}\delta_{2,1'}}{D^2},
 & Q_1\neq Q_2,\\[10pt]
\displaystyle 
\frac{\delta_{1,1'}\delta_{2,2'}+\delta_{1,2'}\delta_{2,1'}}{D^2(1+D_1^{-1})},
& Q_1=Q_2.
\end{cases}
\end{aligned}
\label{eq:Phisecondmoment}
\end{equation}
\end{widetext}
Unlike the first moment, the second moment of $\Phi$ differs from that of the Haar ensemble. 

To quantify this deviation, we compute the trace distance $\Delta_2$ between $\rho_\Phi^{(k=2)}$ and $\rho_{\rm Haar}^{(k=2)}$, defined as $\Delta_2 = \frac{1}{2}{\rm Tr}[\sqrt{\delta\rho_2^\dagger\delta\rho_2}]$, with $\delta \rho_2 = \rho_\Phi^{(k=2)} -\rho_{\rm Haar}^{(k=2)}$. The trace distance measures the maximal probability of distinguishing them using a single optimal measurement. We find that the trace distance (see details in Appendix~\ref{app:analytics}) is exactly given by
\be 
\Delta_2 = \frac{1}{D^2(D+1)}\left[D^2-\sum_{S=0}^{N/2}(2S+1)D_S^2\right].
\ee
After evaluating the combinatorics of $D_S$, we find to leading order in $L$:
\be
\Delta_2 \approx \frac{1}{D}\left(1-\frac{2}{\pi L}\right).
\label{eq:TraceDistance}
\ee
Thus, the trace distance between $\Phi$ and the Haar ensemble is exponentially small in the number of degrees of freedom. It follows that the second moments of observables, including nonlocal ones, coincide with those of Haar-random states up to exponentially small corrections.

More generally, for finite moments ($k\ll D$), one expects the same exponential suppression to persist. Consequently, for any finite-order probe, even those involving highly nonlocal observables, the ensemble $\Phi$ becomes exponentially close to the Haar ensemble, despite being restricted to symmetry-resolved subspaces. 

\subsection{Unentangled Initial States}

We now turn our attention to the initial conditions and show that, although non-Abelian symmetries alone do not preclude states from appearing Haar-random at the level of finite statistical moments, unentangled initial states cannot satisfy Eq.~(\ref{eq:condition}). 
As a result, time-evolved states cannot attain Haar-level randomness, even when the initial state is `infinite-temperature' and has zero total magnetization.

More specifically, we consider product states of $L$ spin-1/2 degrees of freedom, 
\be
|\psi_0\rangle=\bigotimes_{j=1}^L\left(\cos{\frac{\theta_j}{2}}|0\rangle+e^{i\phi_j}\sin{\frac{\theta_j}{2}}|1\rangle\right),
\ee
where each is parametrized in polar coordinates by the angles  $(\theta_j,\phi_j)$ on the Bloch sphere. We restrict the initial conditions to have zero total magnetization, such that $\langle \psi_0 | S_\alpha|\psi_0\rangle = {\rm Tr}[S_\alpha] = 0$, therefore agreeing with the mean value of magnetization of Haar random states. This enforces the condition 
\be
\sum_{i=1}^L (\sin\theta_i\cos\phi_i,
\sin\theta_i\sin\phi_i,
\cos\theta_i)
= \mathbf 0.
\ee

Importantly, the variance $\sigma_\alpha^2 = \langle S_\alpha^2\rangle_\psi - \langle S_\alpha\rangle_\psi^2$ of any product state with zero total magnetization cannot be made arbitrarily large. In particular, all such states satisfy
\be
{\rm Unentangled:} \quad \sigma_x^2+\sigma_y^2+\sigma_z^2 = \frac{L}{2}.
\label{eq:varianceSU2}
\ee
By contrast, Haar-random states have 
\be
{\rm Haar:} \quad \sigma_x^2 + \sigma_y^2 + \sigma_z^2 = \frac{3L}{4},
\ee 
as each spin component satisfies $\langle S_\alpha^2\rangle_{\rm Haar}=\frac{1}{D}{\rm Tr}[S_\alpha^2] = \frac{L}{4}$. This indicates that, for unentangled initial states, the variance of at least one component of the total spin operator must be smaller than that in Haar random states. 

Given the constraint in Eq.~(\ref{eq:varianceSU2}), it is therefore natural to parametrize the initial conditions by two variances, e.g., $\sigma_x^2$ and $\sigma_y^2$, with the third fixed by Eq.~(\ref{eq:varianceSU2}) (always keeping zero total magnetization). This is illustrated in Fig.~\ref{fig:schematics}, where the large triangle bounded by dotted lines indicates the constraint in Eq.~(\ref{eq:varianceSU2}). In addition, the variance of each component of the total spin operator is bounded by $\sigma_\alpha^2 \le L/4$, with $\alpha = x,y,z$, giving rise to the shaded (gray) region enclosed by dashed lines where all physical initial states live. 

We will sample states within the high-symmetry lines labeled $a$--$b$--$c$, while the remaining regions are related by rotations. For instance, $\sigma_x^2=\sigma_y^2=L/2$, $\sigma_z^2=0$ corresponds to states with definite value of magnetization in the $z$ direction, i.e., an `Ising' initial condition with half the spins up and half down along $z$, while $\sigma_x^2=\sigma_y^2=\sigma_z^2=L/6$ corresponds to spins uniformly distributed on the Bloch sphere (`IsoVar' states). As we will see below, 
the late-time entanglement properties of product states are primarily organized by the values $(\sigma_x,\sigma_y,\sigma_z)$. 

\subsection{Bound on Entanglement Entropy}
\label{sec:EEtheory}

Having established that product-state initial conditions cannot satisfy the condition~(\ref{eq:condition}), we now turn to the more quantitative question of how much randomness can be generated through quantum-chaotic dynamical evolution under SU(2) symmetry. We employ entanglement entropy (EE),
\be
S_A = -\text{Tr}[\rho_A \log \rho_A],
\label{eq:EE}
\ee
as a diagnostic of randomness, focusing on the mean EE and, in particular, the subleading corrections to the volume-law scaling (we also comment on fluctuations, but the mean EE already captures the main effect). In Eq.~(\ref{eq:EE}), the reduced density matrix of a given quantum state $|\Psi\rangle$ is defined as $\rho_A = {\rm Tr}_B[|\Psi \rangle\langle\Psi|]$, where the system is bipartitioned into two subsystems of sizes $L_A = f L$ and $L_B=(1-f)L$. 

Because the EE is a nonlinear function of $\rho_A$, it provides an extremely sensitive probe of randomness at high orders—although this same property also makes it difficult, but not impossible~\cite{2019Science_EEmbl,2019Science_renyiions,2021PRL_randomquantumfisher,2023NatRev_randommeasurement,Joshi2023}, to measure experimentally. For our purposes, however, it serves as a powerful theoretical tool to bound the degree of Haar-like randomness that can be generated and, as mentioned above, to sharply diagnose indistinguishability between ensembles beyond local, averaged observables: if the EE across subsystem partitions is insensitive to the distinction between Haar-random states and symmetry-constrained time-evolved states, then one expects other finite-resolution observables to be similarly insensitive.

In the absence of any structure, the distribution of EE for pure random states depends only on the subsystem dimensions through the parameters $f$ and $L$. Without loss of generality, we assume $f \le 1/2$. The average entanglement entropy in the asymptotic limit is given by
\be
\langle S_A \rangle_{\rm Haar} = L f \log d - \frac{1}{2}\delta_{f,1/2},
\label{eq:page}
\ee
a result first conjectured by Page~\cite{1993PRL_Page} and later proven rigorously by others~\cite{2016PRE_entanglementdispersion,2017PRE_entanglementvariance_proof}. The first term on the right-hand side of Eq.~(\ref{eq:page}) is the volume-law contribution, scaling with the subsystem size $L_A = fL$, while the second term produces the characteristic “half-qubit” correction for half-system partitions.
We note that exact expressions exist for both $ \langle S_A \rangle_{\rm Haar}$ and the fluctuations of $S_A$ in {\it finite-size} systems~\cite{2016PRE_entanglementdispersion,2017PRE_entanglementvariance_proof,2019PRD_BianchiDona}. Although we quote here the simple asymptotic forms, all numerical results presented in Sec.~\ref{sec:RQCs} and Sec.~\ref{sec:Ham} use the exact expressions rather than their asymptotic approximations.

We now turn to systems with a single conservation law, such as magnetization along the $Z$ direction. In this case, the EE can be computed exactly when the random states lie entirely within a single symmetry sector, i.e., $\sigma_z^2 = 0$. In this case, the reduced density matrix becomes block diagonal, with block sizes $d_A(M_A) = \binom{L_A}{M_A}\binom{L-L_A}{M-M_A}$, since the number of spin flips in subsystem $A$, denoted $M_A$, is correlated with that in subsystem $B$, denoted as $M_B$, by total magnetization conservation, $M = M_A+M_B$. For random states with well-resolved magnetization $M/L=1/2$, the asymptotic mean EE is given by ~\cite{2017PRL_EEchaotic,2019PRD_BianchiDona,2022PRX_eereview}
\be
\langle S_A \rangle_{\rm U(1)} = \langle S_A \rangle_{\rm Haar} + \frac{f+\log(1-f)}{2}.
\label{eq:meanu1}
\ee
In addition to the volume-law term and the half-qubit shift arising in Eq.~(\ref{eq:page}), Eq.~(\ref{eq:meanu1}) also exhibits a finite offset in the mean EE relative to the Haar result in Eq.~(\ref{eq:page}). We emphasize that this offset is negative (reflecting that these states exhibit less entropy than Haar-random states), is finite for all subsystem fractions $f$, and does {\it not} vanish in the thermodynamic limit. Remarkably, although this result was derived exactly for systems with a simple U(1) scalar charge, it was found that Eq.~(\ref{eq:meanu1}) also describes eigenstates of strongly quantum-chaotic Hamiltonians (with energy in a spatially local Hamiltonian playing the role of the U(1) scalar charge) with remarkably high accuracy~\cite{Huang2021,2024PRX_quantifyingchaos}, even at low energies~\cite{langlett2025quantumchaosfinitetemperature}. The asymptotic EE generalizes straightforwardly to systems with multiple resolved charges, with the O(1) correction increasing proportionally to the number of charges~\cite{2025PRL_EEU1,PhysRevB.108.245101}. 

In the opposite limit, when the constrained states have maximal variance $ \sigma_z^2 = \langle S_z^2 \rangle = L/4 $, they reproduce the statistical properties of Haar-random states for all finite moments~\cite{2025PRB_latetimeensemble}. Although this correspondence can be established analytically only for systems with a simple local U(1) charge, we also found that the conclusion extends to generic quantum-chaotic systems: numerically, the EE of time-evolved states agrees with that of Haar-random states to remarkably high precision for all accessible finite system sizes~\cite{2025PRB_latetimeensemble,2025arxiv_randomizationtimes}. 

\begin{figure}
    \centering
    \includegraphics[width=\linewidth]{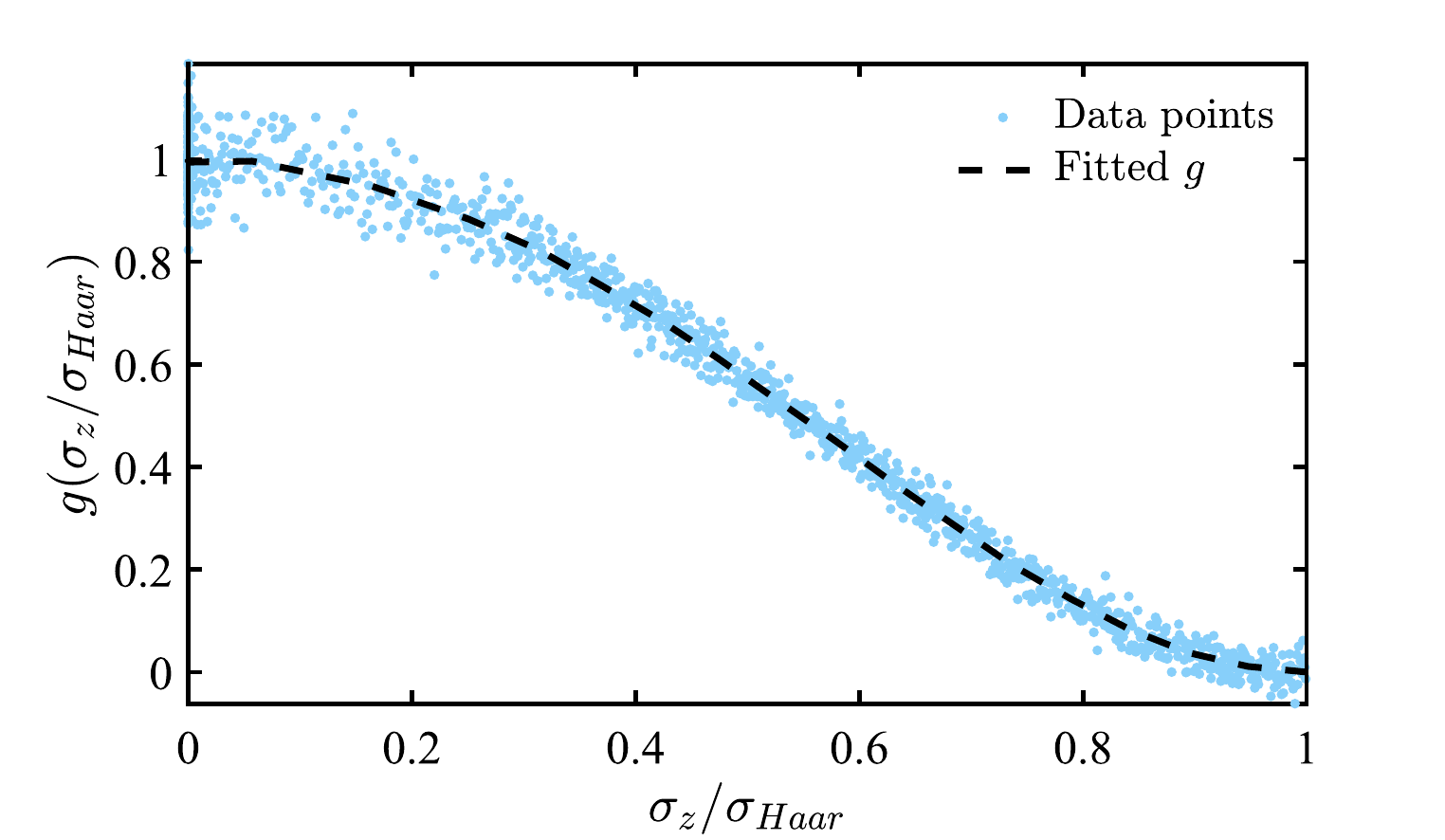}
    \caption{Numerical evaluation of the scaling function $g(x)$ in Eq.~(\ref{eq:EEU1withVar}) as a function of the variance of the conserved charge. We sample $10^3$ states with different system and subsystem sizes, and different values of $\sigma_z$, and find excellent collapse of the datapoints according to Eq.~(\ref{eq:EEU1withVar}).
}
    \label{fig:CorrectionvsVar}
\end{figure}

For intermediate cases where $0 <\sigma_z^2 < \sigma_{\rm Haar}^2 = \frac{{L}}{4}$, the EE interpolates between Eq.~(\ref{eq:page}) and Eq.~(\ref{eq:meanu1}), i.e.,
\begin{equation}
\langle S_A \rangle_{\sigma_z} = \langle S_A \rangle_{\rm Haar} + \frac{f+\log(1-f)}{2}g\left( \frac{\sigma_z}{\sigma_{\rm Haar}}\right), 
\label{eq:EEU1withVar}
\end{equation}
where the dimensionless function $g$ satisfies $g(0)=1$ and $g(1)=0$. While we were unable to find an exact analytical expression for $g$---since the reduced density matrix becomes fully coupled and dense, preventing a direct resolution of its eigenvalues---we determine $g(x)$ numerically by sampling the EE of otherwise random states with zero magnetization and tunable variance $\sigma_z^2$, across different system sizes $L$ and subsystem fractions $f$. As shown in Fig.~\ref{fig:CorrectionvsVar}, we find that Eq.~(\ref{eq:EEU1withVar}) describes the numerical data remarkably well across all system and subsystem sizes and for all values of $\sigma_z$, highlighting the universal character of $g(x)$.

Turning to systems with SU(2) symmetry, we conjecture that Eq.~(\ref{eq:EEU1withVar}) remains valid for each charge component independently, while the noncommutativity of the charges enters only through the constraint between variances in Eq.~(\ref{eq:varianceSU2}), leading to
\begin{equation}
\langle S_A\rangle_{\sigma_x\sigma_y\sigma_z}= \langle S_A\rangle_{\rm Haar}+\sum_{\alpha = x,y,z}\frac{f+\log(1-f)}{2} g\left( \frac{\sigma_\alpha}{\sigma_{\rm Haar}} \right).
\label{eq:EESU2}
\end{equation}
This expectation is motivated in particular by the large-subsystem limit: the typical total spin within subsystem $A$ (among states with zero global magnetization) is extensive, rendering the noncommutativity of the charges a subleading effect in $1/L_A$~\cite{PhysRevB.108.245101}. In this regime, the three components effectively behave as commuting conserved quantities, so that Eq.~(\ref{eq:EEU1withVar}) applies independently to each, with the non-Abelian structure manifest only through the global constraint in Eq.~(\ref{eq:varianceSU2}). Consequently, we expect Eq.~(\ref{eq:EESU2}) to become asymptotically correct in the thermodynamic limit for any finite $f$. Remarkably, we find that even for finite system sizes and relatively small subsystems, Eq.~(\ref{eq:EESU2}) already provides an extremely accurate description of the data, as we will see in Sec.~\ref{sec:RQCs}. 

A consequence of Eq.~(\ref{eq:EESU2}), together with the convexity of $g(x)$ near $x=0$ observed in Fig.~\ref{fig:CorrectionvsVar}, is that the EE is maximized when the variance is distributed as evenly as possible among the three spin directions. In particular, spreading the fluctuations such that $ \sigma_x^2=\sigma_y^2=\sigma_z^2=L/6 $ yields a smaller O(1) correction, $ \delta S_A \approx 0.33\,\frac{f+\log (1-f)}{2} $, than concentrating the variance in two directions while fully resolving one charge, $ \sigma_x^2=\sigma_y^2=L/4,\ \sigma_z^2=0 $, which instead gives $ \delta S_A \approx \frac{f+\log (1-f)}{2} $ (see Fig.~\ref{fig:schematics}(b)).

\section{Random Quantum Circuits constrained by SU(2) symmetry}
\label{sec:RQCs}

\subsection{Setup}

We consider a minimally-structured model of quantum chaotic dynamics comprised of $L$ spin-$\tfrac{1}{2}$ degrees of freedom interacting locally via a RQC model that preserves SU(2) symmetry. We employ the standard brickwork architecture~\cite{2017PRX_quantumcircuits,2023AnnRev_randomcircuits} composed of local two-qubit gates applied in a staggered pattern on even–odd and odd–even bonds, though we note that our results are insensitive to the interaction range of the local gates. 

More specifically, 
at each unit of time $t$ we apply two-qubit gates acting on even-odd sites, followed by two-qubit gates acting on odd-even sites, i.e.,
\begin{equation}
  U_{\text{even}}(t)=\prod_{i\text{ even}}U_{i,i+1}(t),\,\,\,  U_{\text{odd}}(t)=\prod_{i \text{{ odd}}}U_{i,i+1}(t), 
\end{equation}
such that the total unitary evolution at time $T$ is 
\begin{equation}
    \mathcal{U}(T)=\prod_{t=1}^T U_\text{odd}(t)U_\text{even}(t).
\end{equation}
Gates are chosen randomly both in space and in time. 

To enforce SU(2) symmetry in the circuit, we use the local two-site gates 
\begin{equation}
U_{i,i+1}(t)=\exp[i\theta_i(t) \vec{S}_i\cdot\vec{S}_{i+1}],
\end{equation}
where the angles $\theta_i$ are independent random variables uniformly distributed in $\theta_i\in[0,2\pi)$.

We initialize the circuit using product states with zero total magnetization and tunable variances $\sigma_x^2$, $\sigma_y^2$, and $\sigma_z^2$ constrained by the condition in Eq.~(\ref{eq:varianceSU2}). Apart from these constraints, individual spins are otherwise randomly oriented on the Bloch sphere, giving us access to a large sampling space to probe the late-time statistics of quantum states. We generate such states using an iterative algorithm that samples uniformly over all product states consistent with the specified mean and variance; the procedure is detailed in Appendix~\ref{app:initialization}.

The late-time behavior is obtained by evolving each state up to a time $T_0$ sufficient for equilibration (we find $T_0 = 100$ adequate for all accessible system sizes). After equilibration, we construct an ensemble by sampling both over initial conditions and over time, yielding a late-time ensemble 
\be
\Phi_{T\rightarrow\infty}(\sigma_x,\sigma_y,\sigma_z) = \left\{|\Psi_{m,n}\rangle = {\cal U}\left(T_0 + m\right) |\psi_n\rangle\right\},
\ee
where $m$ indexes temporal sampling and $n$ indexes initial-condition sampling. This procedure allows us to compare the state-to-state variability of EE obtained from temporal sampling with that obtained from sampling over initial states at fixed time. As shown in the Appendix~\ref{app:initialvstemporal}, the two are comparable, suggesting that the results appear largely insensitive to the specific choice of $T_0$ and are well organized by the mean and variance of magnetization.

\subsection{Numerical Results}

Returning to Fig.~\ref{fig:schematics}(b), the half-system ($f=1/2$) EE of time-evolved states exhibits large variability across unentangled initial states with zero net magnetization 
and total variance $\sigma_x^2 + \sigma_y^2 + \sigma_z^2 = L/2$. The EE is plotted relative to the Page entropy [Eq.~(\ref{eq:page})], and the shaded region denotes the variance of $S_A$ between samples, obtained by sampling over both initial conditions and evolution times (which yield comparable fluctuations). The progressively darker shading indicates increasing system size $L$. 

Importantly, we find that, regardless of the initial condition, no unentangled initial state approaches the Page entropy characteristic of pure random states, shown as the narrow dark gray band at $\delta S_A = 0$ (the center of the dark gray band indicates the Page entropy, whereas the width indicates the state-to-state fluctuations). This reference is computed using the exact finite-size expressions for the EE and its variance derived in Ref.~\cite{2016PRE_entanglementdispersion,2017PRE_entanglementvariance_proof}, since Eq.~(\ref{eq:page}) gives only the asymptotic thermodynamic-limit result; subleading $O(1/L)$ corrections are comparable to the scale of EE fluctuations for the finite-size numerics discussed here. With increasing $L$, the variance decreases rapidly (indeed, exponentially), while the mean remains approximately constant and systematically below the maximal value. This is consistent with an EE correction that remains finite in the thermodynamic limit. 

Focusing on the families of initial conditions at point $a$ in Fig.~\ref{fig:schematics}(a), having $\sigma_x^2 = \sigma_y^2 = L/4$ and $\sigma_z^2 = 0$, we find that the EE agrees with that of Haar-random states constrained to the largest microcanonical sector of a single U(1) scalar charge (shown as a light-gray horizontal band centered at $\delta S_A \approx -0.09$, whose width is set by the state-to-state fluctuations). As for the Page entropy, we use the exact finite-size expressions for the EE and its variance derived in Ref.~\cite{2019PRD_BianchiDona}, since $O(1/L)$ corrections are comparable to the scale of EE fluctuations in our finite-sized numerics. The close agreement between the SU(2)- and U(1)-constrained ensembles for Ising-like initial conditions ($a$) at the level of half-system EE suggests that other finite-resolution probes may likewise be insensitive to the distinction between these ensembles. As such, the late-time ensemble behaves as if only a single scalar charge were conserved, with maximal mixing between sectors in the $x$ and $y$ directions effectively washing out the associated symmetry.

Scanning the full parameter space of initial conditions, we find that the maximal EE occurs at a single point, denoted $c$, corresponding to the IsoVar state in which the product states are maximally distributed over the Bloch sphere, with $\sigma_x^2 = \sigma_y^2 = \sigma_z^2 = L/6$. We observe remarkably good agreement between the maximal EE obtained numerically and the scaling prediction in Eq.~(\ref{eq:EEU1withVar}), shown with dashed lines, within the scale of the EE fluctuations of the numerical data.

\begin{figure}
    \centering
    \includegraphics[width=1\linewidth]{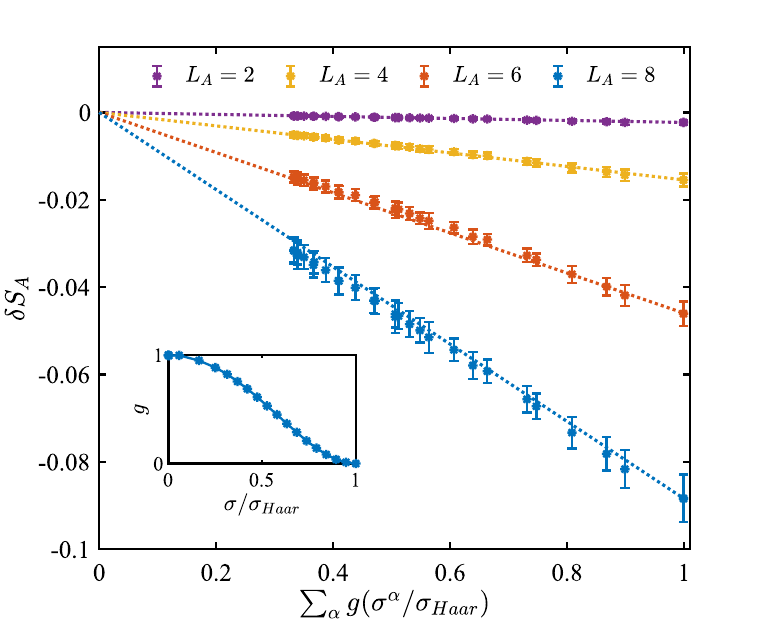}
    \caption{Distribution of EE at late times across all unentangled initial conditions and subsystem sizes. The collapse is shown against the conjectured scaling form of Eq. (\ref{eq:EESU2}), which depends on the variances through the function $\sum_\alpha g(\sigma_\alpha)$ defined therein, and are plotted relative to the Page entropy, $\delta S_A = \langle S_A\rangle - \langle S_A\rangle_{\rm Haar}$. Each point represents the average EE for a given sample, while the error bars indicate the standard deviation obtained from sampling initial states. The dotted lines show the subsystem-size dependence of the function $\frac{f+\log(1-f)}{2}$ appearing in Eq.~(\ref{eq:EESU2}). The data are generated using a random quantum circuit acting on $L=16$ spin-$1/2$ degrees of freedom.}
     \label{fig:EEvsf}
\end{figure}

We now extend the results of Fig.~\ref{fig:schematics} to different subsystem fractions $f$, as shown in Fig.~\ref{fig:EEvsf}. Focusing on system size $L=16$, we collapse all numerical data for the EE correction $\delta S_A$ as a function of the variances $\sigma_x$, $\sigma_y$, and $\sigma_z$, combined into the factor $\sum_\alpha g(\sigma_\alpha/\sigma_{\rm Haar})$ as discussed in connection with Eq.~(\ref{eq:EESU2}). The dataset spans the full accessible space of initial conditions sampled across $(\sigma_x,\sigma_y,\sigma_z)$-space and subsystem fractions $f = 1/8, 1/4, 3/8, 1/2$. We find excellent agreement between the analytical prediction in Eq.~(\ref{eq:EESU2}) and all numerical results, independent of system size, subsystem fraction, and initial condition, supporting the conjecture that Eq. (\ref{eq:EEU1withVar}) remains valid for each charge component independently, while the noncommutativity of the charges enters only through the constraint between variances in Eq.~(\ref{eq:varianceSU2})

\section{Hamiltonian Dynamics constrained by SU(2) symmetry}
\label{sec:Ham}

\subsection{Setup}

We now test whether the mechanism identified in minimally structured RQCs extends to Hamiltonian systems with additional structure, specifically Hamiltonian systems with a global SU(2) symmetry. In this case, there are four conserved charges—$H$, $S_x$, $S_y$, and $S_z$—with the Hamiltonian commuting with the three spin operators:
\be
[H, S_\alpha] = 0, \quad \alpha = x, y, z.
\ee

To construct an SU(2)-invariant Hamiltonian model with the minimal amount of ingredients, we consider a chain of $L$ spin-$1/2$ degrees of freedom with both nearest- and next-nearest-neighbor interactions. The inclusion of next-nearest-neighbor terms is important, as otherwise the Hamiltonian reduces to the integrable Heisenberg model. Specifically, we use
\begin{equation}
H=\sum_{i} \vec{S}_i\cdot\vec{S}_{i+1}
+\lambda_1\vec{S}_i\cdot\vec{S}_{i+2}
+\lambda_2 \vec{S}_i\cdot\bigl(\vec{S}_{i+1}\times\vec{S}_{i+2}\bigr),
\label{eq:Ham}
\end{equation}
where $\vec{S}_i=(S_i^x,S_i^y,S_i^z)$ are the spin-1/2 matrices 
acting on site $i$. This is the most general translationally invariant Hamiltonian with up to three-body interactions built from SU(2)-symmetric spin operators. The first and second terms are the nearest- and next-nearest-neighbor Heisenberg exchange interaction, and the third is the spin-chirality term. We use $\lambda_1=0.9$ and $\lambda_2=0.2$ in our numerics, for which the model exhibits strongly quantum-chaotic behavior, and we impose periodic boundary conditions (we note that the initial condition mixes all momentum sectors).

To initialize the state, we follow the same procedure as that used for RQCs, labeling product-state initial conditions by their variances $\sigma_x^2$, $\sigma_y^2$, and $\sigma_z^2$, with $\sigma_x^2+\sigma_y^2+\sigma_z^2 = L/2$, but with one important addition: because energy is now conserved, we must also impose constraints on the initial energy moments, requiring $E_0 = \langle \psi_0 | H | \psi_0\rangle = {\rm Tr}[H] = 0$, and $\sigma_E^2 = \langle \psi_0 | H^2 | \psi_0\rangle  = \mathrm{Tr}[H^2]$, so that no additional correction to the EE arises from having finite energy density or subtypical energy fluctuations. Although such corrections could be incorporated following the procedure outlined in Sec.~\ref{sec:EEtheory}, our main goal here is to do a direct comparison with the RQC results. 

In the numerical results below, we focus on the two extremal cases: point $a$ in Fig.~\ref{fig:schematics}, corresponding to Ising initial conditions, and point $c$, corresponding to the IsoVar. We employ the same sampling procedure as for the RQCs, retaining only those initial states that have $\langle \psi_0 | H | \psi_0\rangle =0$ and $\langle \psi_0 | H^2 | \psi_0\rangle = \mathrm{Tr}[H^2]$ within a 5\% tolerance. 
Lowering this tolerance does not quantitatively affect the conclusions. Importantly, we find that the inclusion of the $\tau_3$ term in the Hamiltonian is essential; without it, all unentangled initial states will necessarily have subtypical energy variance, $\langle \psi_0 | H^2 | \psi_0\rangle < \frac{1}{D}{\rm Tr}[H^2]$

The procedure for generating the late-time ensemble is analogous to that used for random SU(2) circuits, replacing ${\cal U}(T)$ with the Hamiltonian evolution operator ${\cal U}(t)=e^{-iHt}$. We sample over the initial states defined above and over discrete times $t = T_0 + m\Delta t$, with $T_0 =100$ and $\Delta t = 1$, using the same number of states as in the circuit case.

\begin{figure}
    \centering
    \includegraphics[width=\linewidth]{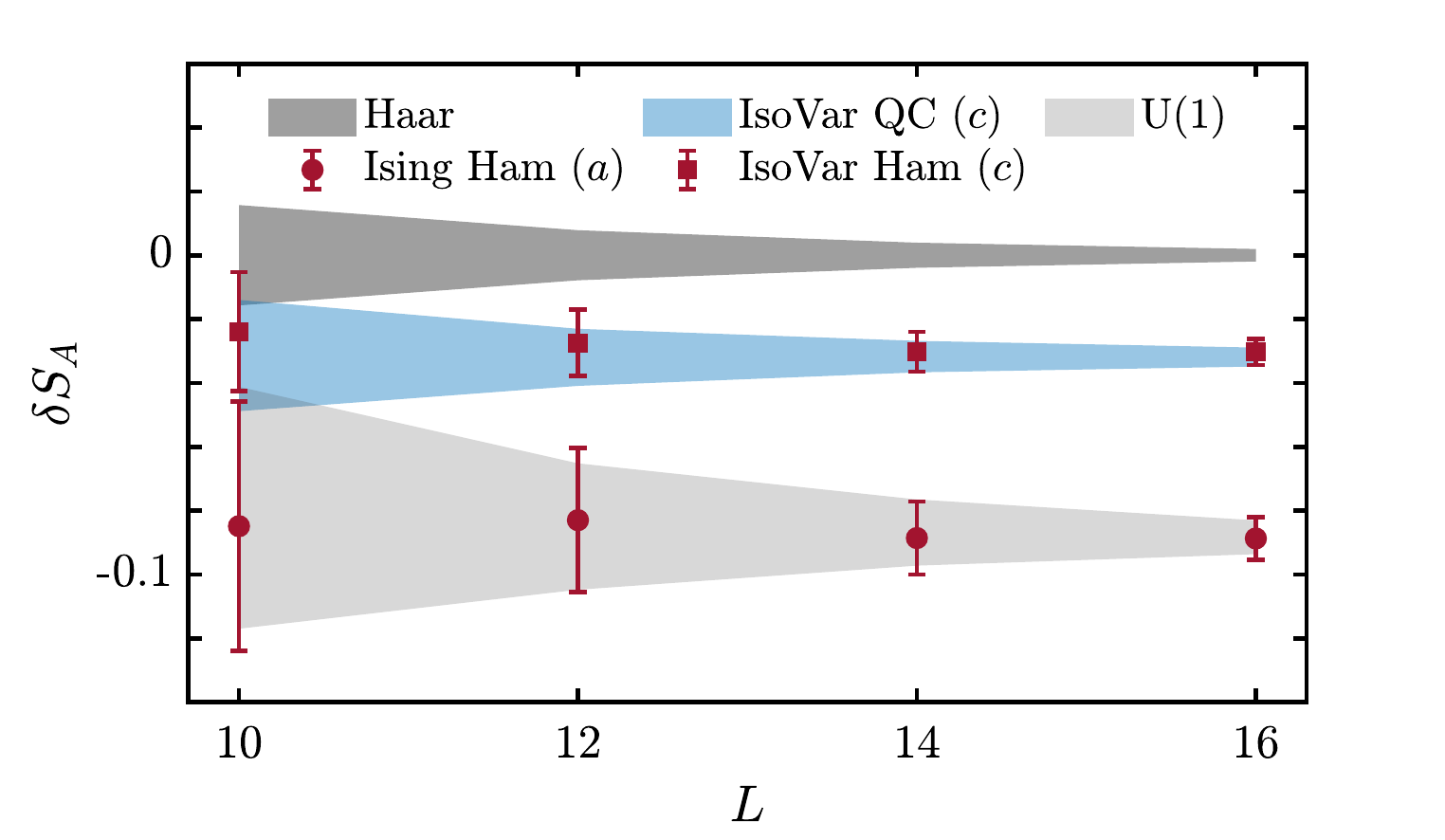}
    \caption{Finite-size scaling of the late-time EE distribution obtained from time evolution of unentangled initial states under the SU(2)-symmetric Hamiltonian in Eq.~\eqref{eq:Ham}. Data are shown relative to the Page entropy, $\delta S_A = \langle S_A\rangle - \langle S_A\rangle_{\mathrm{Haar}}$, for the Ising (circles) and IsoVar (squares) initial conditions. The data points represent the average EE, and the error bars indicate the standard deviation obtained by sampling over initial conditions and evolution times. For comparison, shaded regions show the EE distributions for Haar-random states (dark gray), random states constrained by a single U(1) scalar charge (light gray), and random states constrained by $\sigma_x^2 = \sigma_y^2 = \sigma_z^2 = L/6$ (light blue).}
    \label{fig:Ham}
\end{figure}

\subsection{Numerical Results}

The numerical results for the late-time behavior of product states evolved under the Hamiltonian in Eq.~(\ref{eq:Ham}) are shown in Fig.~\ref{fig:Ham}. As before, the EE is plotted relative to the Page entropy, with points showing the Hamiltonian data for half-system bipartitions as a function of system size. Specifically, the data points represent the average EE for Ising-like initial conditions (point $a$, $\sigma_x^2=\sigma_y^2 = L/2$, $\sigma_z^2=0$; circles) and for the IsoVar initial conditions (point $c$, $\sigma_x^2=\sigma_y^2=\sigma_z^2 = L/6$; squares). The error bars indicate the standard deviation of $\delta S_A$ across initial states and evolution times.

For comparison, we show shaded regions corresponding to the EE distributions for three reference ensembles: pure (Haar) random states (dark gray, centered at $\delta S_A = 0$), pure random states constrained by a single U(1) charge within the largest symmetry sector $\sigma_z = 0$ (light gray), and the late-time ensemble constrained by $\sigma_x^2 = \sigma_y^2 = \sigma_z^2 = L/6$. As before, we use the exact analytical expressions for the Haar and U(1)-constrained cases, including their statistical fluctuations. {Although the shaded regions are shown as continuous bands, the EE is defined only for integer values of $L$; the shading simply interpolates between discrete data points for visual clarity.}

For all system sizes and for both initial conditions, we observe remarkable agreement---not only in the mean EE but also in the scale of its fluctuations, which become exponentially small at the largest system sizes. These results support one of the general conclusions of our work: Although symmetries constrain the exploration of Hilbert space, their effect on randomization dynamics at the level of finite statistical moments is washed out when the initial state is maximally spread across symmetry sectors. This is illustrated in Fig.~\ref{fig:Ham} for the case of energy conservation, which does not contribute an additional correction to the EE, even though it strongly constraints Hilbert space exploration. In contrast, for non-commuting spin operators, such maximal spreading across symmetry sectors is not possible when the initial state is unentangled, and therefore full randomness can never be achieved. 

\section{Discussion}

Our work identified a mechanism that limits the late-time randomization in quantum-chaotic dynamics with non-Abelian symmetries arising from the interplay between symmetry constraints and constraints on quantum-state initialization. At the level of coarse observables, the late-time behavior of quantum states remains well described by conventional statistical mechanics, with local observables and few-body reduced density matrices consistent with thermal ensembles. At the level of fine-grained probes such as entanglement entropy and its fluctuations, however, the late-time ensemble retains memory of higher moments of the conserved charges inherited from the initial state. Incorporating these moments into the statistical description is sufficient to account for the observed late-time entanglement statistics. We emphasize that this does not contradict conventional thermalization; rather, it shows that thermal ensembles correctly describe coarse observables while missing finer statistical information that becomes visible in higher moments and entanglement diagnostics.

For SU(2)-symmetric dynamics, this mechanism has a sharp consequence for experimentally relevant product-state initial conditions. Although chaotic evolution generates highly entangled states, product states cannot reproduce the full Haar-level fluctuations of the conserved spin components, even at the average level. Consequently, their late-time entanglement entropy remains separated from the Page value by an $O(1)$ correction that survives in the thermodynamic limit. Among product states, this correction is smallest for isotropically distributed states.

Looking forward, it would be interesting to investigate whether these ideas extend to other classes of constraints commonly encountered in physical models, such as kinematic constraints or gauge symmetries, which are readily implementable in programmable quantum systems such as Rydberg atom arrays through the Rydberg blockade mechanism.

Separately, a natural question is whether one can implement a simple preparation protocol---e.g., using a finite number of entangling operations---such that the late-time behavior of quantum states matches that of Haar-random states. In this case, it is necessary to relax the SU(2) symmetry conservation during initialization. A naive strategy would be to entangle neighboring spins, for example into singlet pairs between adjacent qubits. Surprisingly, this procedure suppresses rather than enhances the late-time EE, yielding a correction $\delta S_A$ three times larger than that observed for a single scalar charge [see Eq.~(\ref{eq:EEU1withVar})]. This naive example shows that introducing local entanglement during preparation does not necessarily increase entanglement at late times

Finally, an open question concerns the {\it timescales} over which the late-time statistics of constrained random states are approached---specifically, whether the different classes of initial conditions (parametrized by $\sigma_x$, $\sigma_y$, and $\sigma_z$) can exhibit parametrically distinct equilibration scaling with system size. In recent work on systems with one scalar charge, we found---somewhat surprisingly---that certain classes of initial conditions can lead to parametrically faster equilibration with system size, even faster than that suggested by the sub-ballistic growth of entanglement discussed in the context of random-circuit models with a U(1) conservation law. Given that the space of initial conditions is considerably richer in the SU(2) case than in the U(1) case, it would be interesting to investigate the connection between initialization and equilibration timescales, in particular whether one can achieve scaling distinct from ballistic and diffusive behavior.

\section*{Acknowledgements}

JRN acknowledges the hospitality of the Kavli Institute for Theoretical Physics through the program {\it Learning the Fine Structure of Quantum Dynamics in Programmable Quantum Matter}, supported by NSF Grant No.\ PHY-2309135. JRN also acknowledges the hospitality of the Aspen Center for Physics, which is supported by NSF Grant No.\ PHY-2210452 and by a grant from the Alfred P. Sloan Foundation (G-2024-22395). Numerical simulations were performed using the advanced computing resources provided by Texas A\&M High Performance Research Computing.

\appendix

\section{Adding the zero magnetization constraint on the ensemble $\Phi$}
\label{app:fullSU2}

Here we construct an ensemble of states analogous to that in Eq.~(\ref{eq:constrainedensemble}), but with the stronger constraint that each individual state, not just the ensemble average, has vanishing total magnetization, $\langle \Phi_i | S_\alpha | \Phi_i \rangle = 0$ ($\alpha=x,y,z$), while also reproducing the same spin fluctuations $\langle S_\alpha^2\rangle = \frac{1}{D}{\rm Tr}[S_\alpha^2]$. We show numerically that this more constrained ensemble still exhibits the same statistical properties as Haar-random states, up to corrections exponentially small in system size.

To construct such states, we first decompose the Hilbert space into total-spin sectors $S$. For example, for $L=4$ spin-$\tfrac12$ degrees of freedom, $\bigotimes_{i=1}^4 \tfrac12 = 2 \oplus 1 \oplus 1 \oplus 1 \oplus 0 \oplus 0$. Within each spin-$S$ sector, of dimension $2S+1$, we generate a state $|\phi_S\rangle=\sum_m c_m |S,m\rangle$ such that $\langle \phi_S|\vec S|\phi_S\rangle=0$, $\langle S_x^2\rangle=\langle S_y^2\rangle=\langle S_z^2\rangle=\frac{S(S+1)}{3}$. A convenient way to do this is to generate trial states of the form
\be
|\phi_S\rangle=\sum_m \frac{e^{i\theta_m}}{\sqrt{2S+1}}|S,m\rangle,
\ee
with random phases $\theta_m$, and then resample the phases until these constraints are satisfied up to a prescribed numerical tolerance. We then construct a seed state in the full Hilbert space as
\be
|\Phi_0\rangle=\sum_S \sqrt{\frac{D_S}{D}} |\phi_S\rangle,
\ee
where $D_S$ is the multiplicity of spin sector $S$ and $D$ is the total Hilbert-space dimension. This choice ensures that
\be
\langle \psi|\vec S|\psi\rangle=0,
\qquad
\sigma_x^2=\sigma_y^2=\sigma_z^2=L/4.
\ee
Finally, applying a global random SU(2) unitary to $|\Phi_0\rangle$ generates the constrained ensemble 
\be
|\Phi_i\rangle = U_i|\Phi_0\rangle,
\label{eq:constrainedensemble2}
\ee
while preserving the desired constraints.

\begin{figure}
    \includegraphics[width=\linewidth]{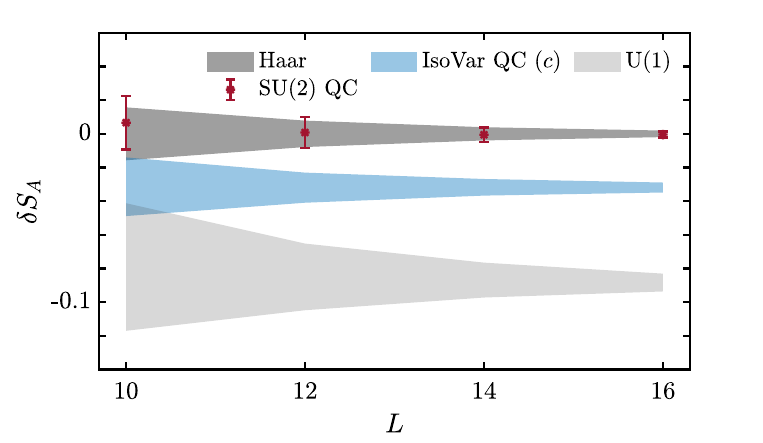}
    \caption{Finite-size scaling of EE distribution for the constrained ensemble defined in Eq.~(\ref{eq:constrainedensemble2}). Data are shown relative to the Page entropy, $\delta S_A = \langle S_A\rangle - \langle S_A\rangle_{\mathrm{Haar}}$. The data points represent the average EE, and the error bars indicate the standard deviation obtained by sampling over states $|\Phi_i\rangle$ in Eq.~\ref{eq:constrainedensemble2}. For comparison, shaded regions show the EE distributions for Haar-random states (dark gray), random states constrained by a single U(1) scalar charge (light gray), and random states constrained by $\sigma_x^2 = \sigma_y^2 = \sigma_z^2 = L/6$ (light blue).}
    \label{fig:randomzeromagnetization}
\end{figure}

We benchmark this ensemble using several diagnostics of randomness. For example, Fig.~\ref{fig:randomzeromagnetization} plot the EE distribution of the constrained ensemble, showing good agreement with Haar-random states not only at the level of the average EE, but also at the level of the fluctuations, which are exponentially small in system size. These results support the conclusions of Sec.~\ref{sec:contrainedensemble}: even when zero total magnetization is imposed on each individual realization, the ensemble still reproduces the finite-resolution statistical properties of Haar-random states when probed through fine-grained diagnostics of randomness. 

\section{Trace distance between Haar random states and typical random states}
\label{app:analytics}

To compute the trace distance between the second moment of the ensemble of Haar-random states, Eq.~(\ref{eq:Haarsecondmoment}), and the second moment of the ensemble $\Phi$, Eq.~(\ref{eq:Phisecondmoment}), we first note that the matrix $\delta \rho^{(k=2)}$ is sparse, with nonzero off-diagonal entries appearing only for permutations of indices with $m_1 \neq m_2$, i.e., $[\delta \rho^{(k=2)}]_{m_1 m_2,, m_2 m_1} \neq 0$. To diagonalize this sparse matrix and compute the trace distance, we work in the basis $|Q,\alpha\rangle$, where $Q$ is a shorthand notation $Q=(M,S)$ labeling the symmetry sectors with total magnetization $M$ associated with the $S_z$ operator and total spin $S$ associated with $S^2=S_x^2+S_y^2+S_z^2$. The index $\alpha \in [1,D_Q]$ labels the state components within each sector, where $D_Q$ is equal to the total Hilbert space with fixed spin number $S$, $ D_S = {L \choose L/2-S} - {L \choose L/2-S-1}$.

Starting from Eq.~(\ref{eq:Phisecondmoment}) in the main text, we need to consider three distinct cases:
\begin{itemize}
\item When $Q_1 \neq Q_2$, there is a total of $ \sum_{Q_1\neq Q_2} D_{Q_1}D_{Q_2}$ terms that contribute to the trace distance. Each state $(Q_1\alpha_1,Q_2\alpha_2)$ is coupled to the state obtained by permuting its indices, $(Q_2\alpha_2,Q_1\alpha_1)$, and the eigenvalues of the two-by-two matrix describing this subspace are 0 and $\frac{1}{D^2}-\frac{1}{D^2(1+1/D)}$. We also note that $\sum_{Q_1\neq Q_2} D_{Q_1}D_{Q_2} = D^2-\sum_QD_Q^2$.

\item When $Q_1 = Q_2 = Q$ and $\alpha_1\neq\alpha_2$, for {\it each} symmetry sector $Q$ there is a total of $D_Q(D_Q-1)$ terms that contribute to the trace distance. Each state $(Q\alpha_1,Q\alpha_2)$ is only coupled to the state obtained by permuting its indeces, $(Q\alpha_2,Q\alpha_1)$, and the eigenvalues of the two-by-two matrix describing this subspace are 0 and $\frac{1}{D^2(1+1/D)}-\frac{1}{D^2(1+1/D_Q)}$. 

\item When $Q_1 = Q_2 = Q$ and $\alpha_1=\alpha_2$, there is a total of $D_Q$ terms that contribute to the trace distance in {\it each} symmetry sector $M$. Each term contributes $\frac{1}{D^2(1+1/D_Q)}-\frac{1}{D^2(1+1/D)}$ to the trace distance. 
\end{itemize}

Combining all three contributions (i)-(iii), the trace distance is given by 
\begin{align}
\Delta_{k=2} & = \frac{1}{2}\left(1-\frac{1}{D^2}\sum_QD_Q^2\right)\left(1-\frac{1}{1+1/D}\right) \nonumber\\
 & + \sum_Q \frac{D_Q(D_Q-1)}{2D^2}\left( \frac{1}{1+1/D} - \frac{1}{1+1/D_Q}\right) \nonumber\\
 & + \sum_Q \frac{D_Q}{D^2} \left(\frac{1}{1+1/D_Q} - \frac{1
}{1+1/D}\right),
 \label{eq:delta2exact}
\end{align}
which, after rearranging all the terms results in:
\be 
\Delta_2 = \frac{2}{D^2(D+1)}\left[D^2-\sum_{S=0}^{N/2}(2S+1)D_S^2\right].
\ee

The final step in the derivation is to evaluate the quantity $\sum_Q D_Q^2=\sum_S (2S+1)D_S^2$. Using the expression for the multiplicities $D_S = {L \choose L/2-S} - {L \choose L/2-S-1}$, the sum can be written as a sum over spin sectors $S$. Introducing the variable $k = L/2 - S$ allows us to rewrite the expression in terms of binomial coefficients of the form ${L \choose k} - {L \choose k-1}$. The resulting sum can then be simplified using the Vandermonde identity, yielding the compact result
\be
\sum_Q D_Q^2 = {N \choose N/2}^2 .
\ee
Finally, applying Stirling’s approximation ${L \choose L/2}^2 \approx \frac{2D^2}{\pi L}$ to the central binomial coefficient gives the asymptotic expression quoted in Eq.~(\ref{eq:TraceDistance}).

\section{Product State Initialization}
\label{app:initialization}

This section explains how to generate random product state initial conditions with a specified variance and mean of $S_z$. A general pure product state can be written as
\begin{equation}
    |\psi\rangle=\bigotimes_{i=1}^L\left(\cos{\frac{\theta_i}{2}}|0\rangle+e^{i\phi_i}\sin{\frac{\theta_i}{2}}|1\rangle\right),
\end{equation}
where $\theta_i\in[0,\pi]$ and $\phi_i\in[0,2\pi]$. Firstly, we want to set the mean value of all three directions of the total spin operator to be zero.
\begin{equation}
    \begin{split}
        \langle S_x\rangle&=\frac{1}{2}\sum_{i=1}^L\sin{\theta_i}\cos{\phi_i}=0,\\
        \langle S_y\rangle&=\frac{1}{2}\sum_{i=1}^L\sin{\theta_i}\sin{\phi_i}=0,\\
        \langle S_z\rangle&=\frac{1}{2}\sum_{i=1}^L\cos{\theta_i}=0.
    \end{split}
\end{equation}
To make the notation simple, we define $x_i=\sin{\theta_i}\cos{\phi_i}$, $y_i=\sin{\theta_i}\sin{\phi_i}$ and $z_i=\cos{\theta_i}$. The above equations can be rewritten as
\begin{equation}
    \sum_{i=1}^L x_i=\sum_{i=1}^L y_i=\sum_{i=1}^L z_i=0.
    \label{eq:zerozum}
\end{equation}
Secondly, we want to set the variance of $S_x$, $S_y$ and $S_z$ to be specific values $\sigma_x^2$, $\sigma_y^2$ and $\sigma_z^2$. The variance of $S_x$, $S_y$ and $S_z$ can be rewritten as
\begin{equation}
    \begin{split}
        \sigma_x^2&=\frac{L}{4}-\frac{1}{4}\sum_{i=1}^L x_i^2,\\
        \sigma_y^2&=\frac{L}{4}-\frac{1}{4}\sum_{i=1}^L y_i^2,\\
        \sigma_z^2&=\frac{L}{4}-\frac{1}{4}\sum_{i=1}^L z_i^2.
    \end{split}
    \label{eq:typicalVar}
\end{equation}
Therefore, the problem becomes finding $3L$ variables $\{x_i,y_i,z_i\}$ that satisfy Eqs.~(\ref{eq:zerozum}) and (\ref{eq:typicalVar}) together with the constraint
\begin{equation}
    x_i^2+y_i^2+z_i^2=1
\end{equation}
for $i=1,2,...,L$. We define the variances in terms of the variable $A_x$, $A_y$ and $A_z$ as
\begin{equation}
    \begin{split}
        A_x = \frac{1}{L}\sum_{i=1}^L x_i^2&=1-4\frac{\sigma_x^2}{L},\\
        A_y = \frac{1}{L}\sum_{i=1}^L y_i^2&=1-4\frac{\sigma_y^2}{L},\\
        A_z = \frac{1}{L}\sum_{i=1}^L z_i^2&=1-4\frac{\sigma_z^2}{L}.
    \end{split}
\end{equation}

To solve these equations, we can use the following algorithm. Firstly, we sample $\{x_i\}$ from the normal distribution $\mathcal{N}(0,\sqrt{A_x})$, $\{y_i\}$ from $\mathcal{N}(0,\sqrt{A_y})$, and $\{z_i\}$ from $\mathcal{N}(0,\sqrt{A_z})$. Secondly, we shift the variables to satisfy the zero mean condition, e.g., $x_i\rightarrow x_i-\frac{1}{L}\sum_{j=1}^L x_j$. Finally, we normalize the variables to satisfy the constraint $x_i^2+y_i^2+z_i^2=1$. Then calculate the value $A_x'=\frac{1}{L}\sum_{i=1}^{L}x_i^2$, and shift the variables again by $x_i\rightarrow x_i\sqrt{A_x/A_x'}$. We repeat the normalization and shifting process for $y_i$ and $z_i$ until all six equations are satisfied within a certain tolerance. Finally, we can calculate $\theta_i$ and $\phi_i$ from $x_i$, $y_i$ and $z_i$ by
\begin{equation}
        \theta_i=\arccos{z_i},\quad    \phi_i=\arctan{\frac{y_i}{x_i}}.
\end{equation}  
With the above algorithm, we generate ensembles of random product states with specified spin variance and zero total magneteization. 

\begin{algorithm}[H]
    \caption{Generate product states with given spin variances}
    \begin{algorithmic}[1]
        \State Choose target variances $\sigma_\alpha^2$ and compute $A_\alpha = 1 - 4\sigma_\alpha^2/L$ for $\alpha \in \{x,y,z\}$
        \State Sample $x_i \sim \mathcal{N}(0,\sqrt{A_x})$, $y_i \sim \mathcal{N}(0,\sqrt{A_y})$, $z_i \sim \mathcal{N}(0,\sqrt{A_z})$
        \State Enforce zero means: $v_i \gets v_i - \frac{1}{L}\sum_j v_j$ for $v \in \{x,y,z\}$
        \Repeat
            \For{$i=1$ to $L$}
                \State $r_i \gets \sqrt{x_i^2 + y_i^2 + z_i^2}$
                \State $(x_i,y_i,z_i) \gets (x_i,y_i,z_i)/r_i$
            \EndFor
            \For{$v \in \{x,y,z\}$}
                \State $A_v' \gets \frac{1}{L}\sum_i v_i^2$
                \State $v_i \gets v_i\sqrt{A_v/A_v'}$
            \EndFor
        \Until{$|A_v' - A_v| < \epsilon$ for all $v$}
        \For{$i=1$ to $L$}
            \State $\theta_i \gets \arccos(z_i)$
            \State $\phi_i \gets \arctan(y_i/x_i)$
        \EndFor
    \end{algorithmic}
\end{algorithm}

\begin{figure}
    \centering
    \includegraphics[width=\linewidth]{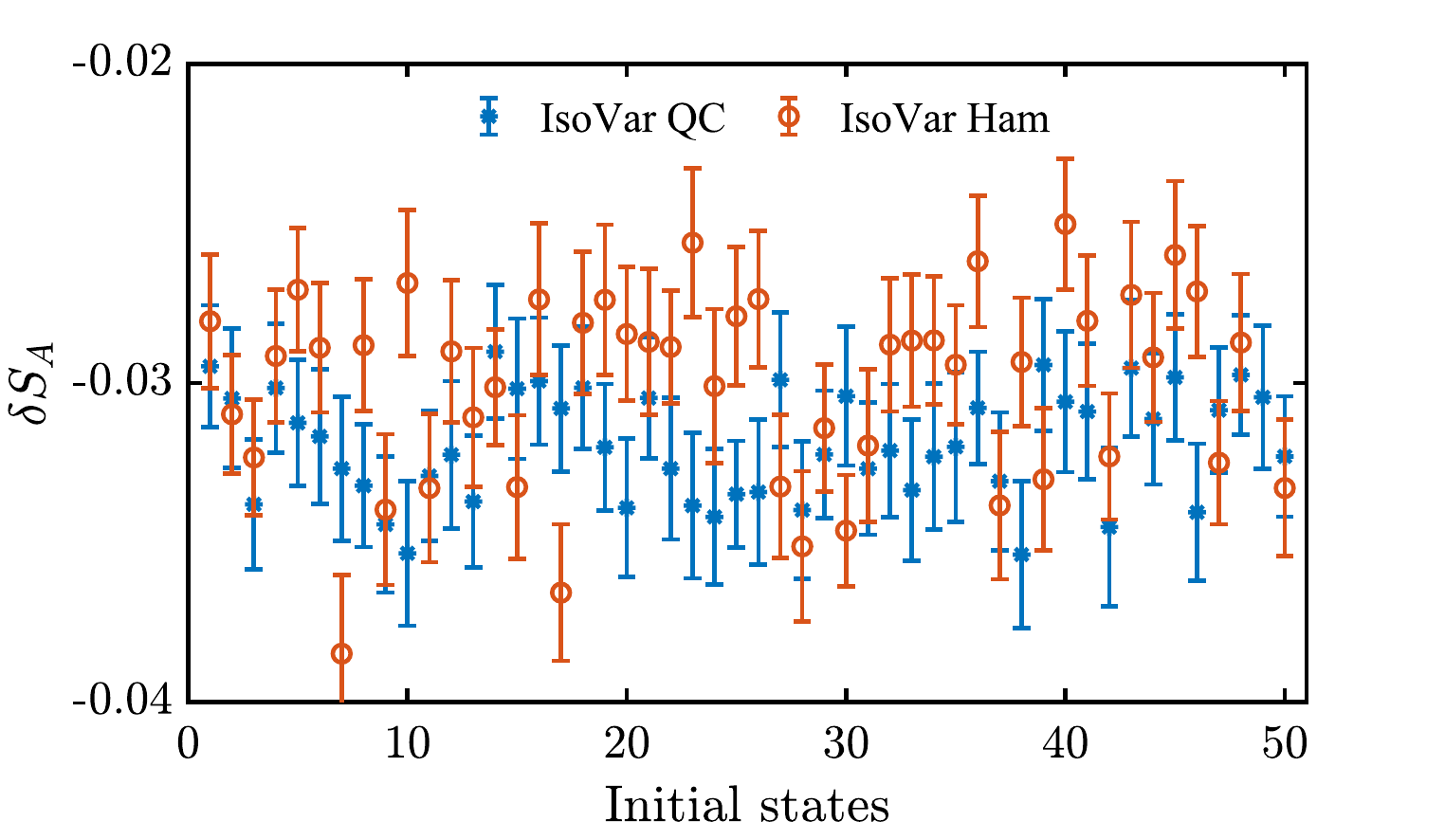}
    \caption{Mean and state-to-state fluctuations of the EE for 50 different initial product states. Each data point represents the time-averaged EE, while the error bars indicate the temporal fluctuations.}
    \label{fig:CompareTemporalSampling}
\end{figure}

\section{Temporal vs sampling variance}
\label{app:initialvstemporal}

In this section we provide additional numerical data comparing state-to-state fluctuations in the EE when sampling over initial conditions (all having zero total magnetization and equal $\sigma_x$, $\sigma_y$, and $\sigma_z$) and when sampling one initial state in time, showing that both are comparable.

For each initial state $|\psi_{n}\rangle$, we evolve the system under unitary evolution constrained by SU(2) symmetry until it equilibrates. We then sample $N$ different time steps $\{t_1,t_2,\ldots,t_N\}$ to construct a late-time ensemble of states:
\be
|\Psi_{m,n}\rangle = {\cal U}(t_m)|\psi_{n}\rangle ,
\ee
where $n$ labels the initial condition and $m$ labels the sampled time. The resulting late-time ensemble is therefore comprised of  $M \times N$ states.

In Fig.~\ref{fig:CompareTemporalSampling}, we show numerical results for product-state initial conditions with $\sigma_x^2=\sigma_y^2=\sigma_z^2=\frac{L}{6}$, evolved under both RQC dynamics and Hamiltonian evolution. We choose $M=50$ initial states and sample $N=200$ time points. The system size is $L=16$ and the subsystem size is $L_A=8$. We find that, provided the initial states share the same magnetization variances, the temporal fluctuations of the entanglement entropy (EE) are comparable to those obtained by sampling over different initial conditions. This suggests that the late-time dynamics is primarily controlled by the total spin variance across the three spin orientations.

 \section{Prepare initial states for Hamiltonian}
\label{app:Haminitial}

Here we extend the procedure described in App.~\ref{app:initialization} to account for energy constraints. In particular, the states produced using the algorithm described in App.~\ref{app:initialization} are typically not located near the middle of the energy spectrum and may also exhibit smaller-than-typical energy fluctuations. These effects introduce additional corrections to the EE, in particular those associated with the energy operator.

\begin{figure}
    \centering
    \includegraphics[width=\linewidth]{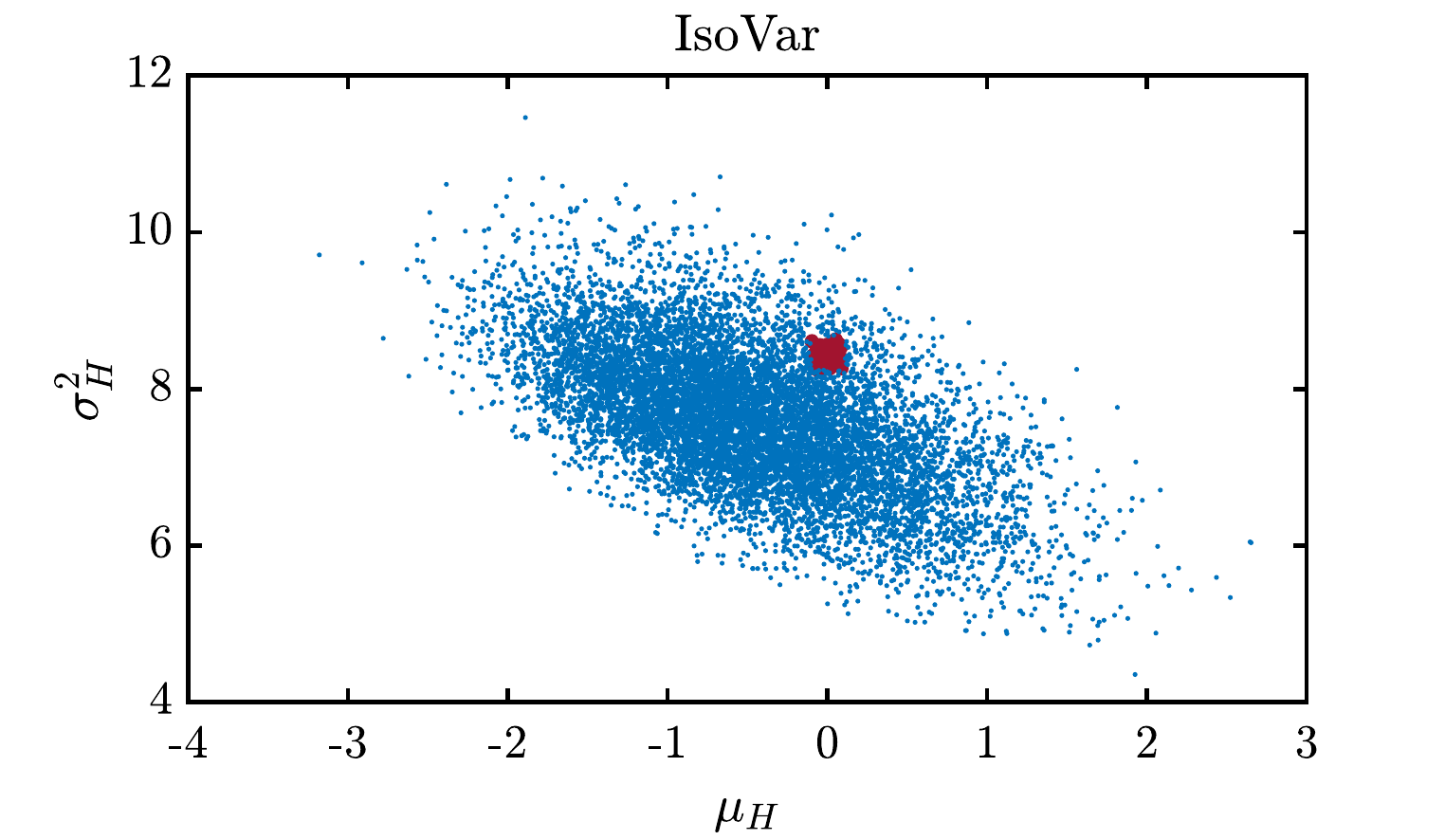}[b]
    \caption{Procedure used to generate initial states for the Hamiltonian described in Sec.~\ref{sec:Ham}. The data points are produced using the algorithm described in App.~\ref{app:initialization}, followed by selecting states whose mean energy and variance lie within 5\% of the target values defined in Eq.~(\ref{eq:Hconditions}).}
    \label{fig:HamInitialState}
\end{figure}

To account for these effects, when preparing initial states for the Hamiltonian discussed in Sec.~\ref{sec:Ham}, we additionally impose the conditions
\be
\langle \psi_0 | H | \psi_0 \rangle = 0, \qquad
\langle \psi_0 | H^2 | \psi_0 \rangle = {\rm Tr}[H^2],
\label{eq:Hconditions}
\ee
within a prescribed tolerance for both quantities. In practice, we set this tolerance to 5\% of the characteristic scale of the fluctuations. To generate such states, we employ a trial-and-error procedure in which candidate states are produced using the algorithm described in App.~\ref{app:initialization} and retained only if they satisfy Eq.~(\ref{eq:Hconditions}) within the specified tolerance. An illustration of this procedure for the case $\sigma_x^2 = \sigma_y^2 = \sigma_z^2 = L/6$ is shown in Fig.~\ref{fig:HamInitialState}, demonstrating that there is a small but finite probability of obtaining samples that satisfy Eq.~(\ref{eq:Hconditions}).

\end{document}